\documentclass[aip,reprint,english,groupedaddress]{revtex4-1}
\usepackage{graphicx}
\usepackage[T1]{fontenc}
\usepackage[latin9]{inputenc}
\usepackage{amstext}
\usepackage{amssymb}
\usepackage{multirow}
\usepackage{dcolumn}
\usepackage{braket}
\usepackage{amsmath}
\usepackage{setspace}
\usepackage{epstopdf}
\usepackage{babel}
\makeatletter

\begin{document}

\title{Optimized Coplanar Waveguide Resonators for a Superconductor--Atom Interface}

\author{M. A. Beck}
\email[]{mabeck2@wisc.edu}
\author{J. A. Isaacs}
\author{D. Booth}
\author{J.D. Pritchard}
\thanks{Current address: Department of Physics, University Of Strathclyde, 107 Rottenrow East, Glasgow, United Kingdom}
\author{M. Saffman}
\author{R. McDermott}
\affiliation{Department of Physics, University Of Wisconsin - Madison, 1150 University Avenue, Madison, Wisconsin 53706}

\date{\today}
\begin{abstract}

We describe the design and characterization of superconducting coplanar waveguide cavities tailored to facilitate strong coupling between superconducting quantum circuits and single trapped Rydberg atoms. For initial superconductor--atom experiments at 4.2~K, we show that resonator quality factors above $10^4$ can be readily achieved. Furthermore, we demonstrate that the incorporation of thick-film copper electrodes at a voltage antinode of the resonator provides a route to enhance the zero-point electric fields of the resonator in a trapping region that is 40 $\mu$m above the chip surface, thereby minimizing chip heating from scattered trap light. The combination of high resonator quality factor and strong electric dipole coupling between the resonator and the atom should make it possible to achieve the strong coupling limit of cavity quantum electrodynamics with this system.
\end{abstract}

\maketitle 

Quantum computers will enable efficient solution of problems that are intractable on conventional, classical computers. A number of candidate physical systems for quantum bits (''qubits'') are currently under investigation, including superconducting integrated circuits incorporating Josephson junctions \cite{Devoret:2013aa,Kelly:2015aa,Corcoles:2015aa}, semiconducting quantum dots \cite{Loss:1998aa, Zwanenburg:2013aa,Kim:2014aa}, trapped neutral atoms \cite{Saffman:2010aa,Specht:2011aa,Bloch:2012aa}, and trapped ions \cite{Blatt:2008aa, Monroe:2014aa}. The various approaches each have strengths and weaknesses, and there are unsolved scientific challenges associated with scaling any of the current technologies. Against this backdrop, there has been a growing interest in the last several years in hybrid approaches to quantum information processing that combine the best features of several different methods \cite{Sorensen:2004aa, Verdu:2009aa, Verdu:2009aa, Schuster:2011aa, Bernon:2013aa, OBrien:2014aa, Thiele:2015aa, Voigt:2015aa, Weiss:2015aa, Kurizki31032015}. Recent efforts to interface disparate quantum systems include coupling superconducting resonators to quantum dots\cite{Frey:2012aa}, electronic spin ensembles\cite{Kubo:2011aa}, and neutral atom clouds\cite{Hermann:2014aa}.

One attractive hybrid approach would involve a fast, high-fidelity superconducting quantum processor coupled to a stable, long-lived neutral atom quantum memory via a Rydberg state. Superconductor gate times are of order 10 ns, and fidelities are now at the threshold for fault-tolerance in the surface code\cite{Fowler:2012aa}; however, coherence times are typically tens of $\mu$s. In contrast, neutral atoms offer coherence times of order seconds, so that the superconductor--atom system would yield an unprecedented ratio of coherence time to gate time.
Moreover, a superconductor--atom quantum interface could open the door to efficient microwave-to-optical photon conversion, an essential ingredient in a distributed quantum information processing network \cite{Cirac:1997aa,Kimble:2008aa}. 

The key technological obstacle to realization of a hybrid superconductor--atom system is the microwave photon--atom interface. Prior attempts to combine trapped neutral atoms with thin-film superconducting cavities have relied on magnetic coupling \cite{Verdu:2009aa, Bernon:2013aa, Voigt:2015aa}; due to the smallness of the magnetic moment, these schemes require coupling to atomic ensembles to achieve appreciable interaction strengths. An alternative approach is to couple the electric dipole moment of a single trapped Rydberg atom to the zero-point electric field of the resonator \cite{Pritchard:2014aa}. As we show below, an appropriately designed superconducting resonator should allow realization of coupling strengths to a single atom in the MHz range, corresponding to the strong coupling limit of cavity quantum electrodynamics (QED). While the ultimate goal is to realize a superconductor--atom interface at millikelvin temperatures, the integration of cold atoms in a millikelvin-temperature cryostat presents formidable technical challenges. Accordingly, we are pursuing the intermediate goal of interfacing a single trapped atom to a $\sim$5 GHz resonator in a 4.2~K liquid helium (LHe) cryostat. Despite the nonnegligible thermal occupation $\bar{n}\sim20$ of the microwave mode, the 4.2~K test bed should still enable detailed study of the spatial dependence of the superconductor--atom vacuum Rabi frequency, the Purcell enhancement of the Rydberg lifetime, and possible deleterious interactions between the atom and the surfaces of the superconducting waveguide structure.
\begin{figure*}[t]\label{fig:aspectData}
\centering
	
	\includegraphics[width= 1.0\textwidth]{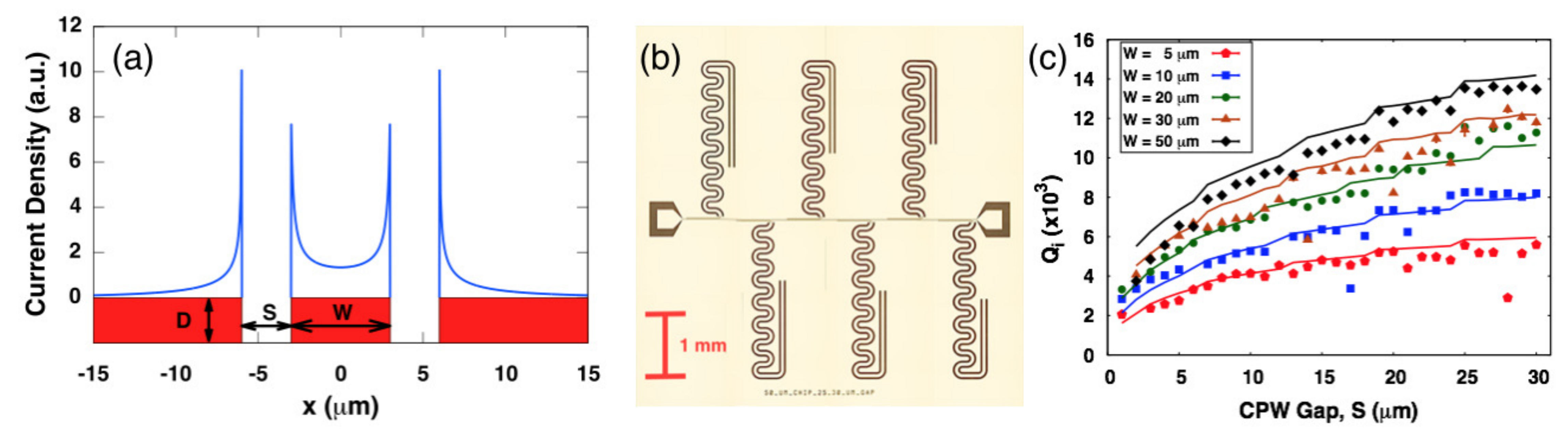} \hfill \\[-3ex]
	
	\caption{(Color Online) (a) Normalized supercurrent density (Blue) in a CPW with center trace width $W = 6 \, \mu$m, gap width $S =  3\, \mu$m, and zero-temperature penetration depth $\lambda_0 = 87$ nm. The values displayed are averaged over the thickness of the traces, $D = 100$ nm. (b) Optical micrograph of multiplexed CPW chip for investigation of dependence of resonator quality factor on geometry. (c) CPW internal quality factor as a function of CPW trace width $W$ and gap width $S$ as measured at 4.2~K. The resonant frequencies covered a span of 4.8-5.2 GHz. Solid lines are theoretical predictions from Eq.~\eqref{Qcpw}. The discontinuities in the theoretical predictions arise from incorporating the slightly different resonator center frequencies in the calculation of the complex conductivity. Error bars for the fits are smaller than the symbol size.}
	\label{Fig1}
\end{figure*}

In this Letter, we describe the design and characterization of superconducting coplanar waveguide (CPW) resonators tailored to facilitate strong coupling to single trapped Rydberg atoms at a temperature of 4.2~K. Our approach minimizes loss in the resonator due to thermal quasiparticle excitations in the superconductor, leading to long lifetimes for the microwave cavity photons. In addition, our design provides enhanced zero-point electric fields at a region that is remote from the surface of the superconducting chip, thereby minimizing chip heating due to scattered trap light. We show that the combination of low loss and large zero-point fields firmly places the superconductor--atom interaction in the strong coupling limit of cavity QED.

At temperatures approaching the superconducting transition temperature $T_c$, thermal quasiparticles represent the dominant source of microwave loss. The quasiparticle-limited quality factor $Q_{\text{QP}}$ in a superconducting resonator is written as
\begin{align} \label{Qcpw}
Q_{\text{QP}} =  \frac{\omega_c (L_k + L_g)}{R_s} = \frac{\sigma_2}{\sigma_1}\left(1 + \frac{L_g}{L_k}\right) ,
\end{align}
where $R_s+iL_k$ is the complex surface impedance of the superconductor, with kinetic inductance per unit length $L_k$; $L_g$ is the geometric inductance per unit length of the resonator; and $\sigma = \sigma_1 - i\sigma_2$ is the temperature- and frequency-dependent complex conductivity of the superconductor \cite{Mattis:1958aa}. For the CPW geometry, the geometric inductance per unit length is given by $L_g = \mu_0 K(k')/4K(k)$ \cite{Clem:2013aa}, where $K$ is the complete elliptic integral of the first kind; $k = W/(W+2S)$, $k' = \sqrt{1 - k^2}$; and $W$ and $S$ are the CPW center trace and gap width, respectively [see Fig.~\ref{Fig1}(a)]. The kinetic inductance per unit length $L_k$ is defined via the kinetic energy of the supercurrent as follows \cite{Tinkham}:
\begin{align}\label{Lk}
E_k \equiv \frac{1}{2} L_k I^2 = \frac{1}{2} \mu_0 \lambda^2 \int j^2 dA,
\end{align}
where $\lambda$ is the superconducting penetration depth. For trace thicknesses $D \ll \lambda$, the current density is approximately uniform over the cross sectional area of the CPW and Eq.~\eqref{Lk} can be evaluated analytically. In general, however, the current density is highly non-uniform \cite{VanDuzer} with the highest density at the trace edges, necessitating numerical evaluation of Eq.~\eqref{Lk}. Fig.~\ref{Fig1}(a) displays the normalized current density for a CPW with trace width $W = 6\, \mu$m and gap width $S = 3\, \mu$m. Qualitatively, in the limit $S \gg W$, the geometric contribution to the inductance reduces to $L_g \rightarrow \mu_0$, while the kinetic contribution reduces to $L_k \rightarrow \mu_0 \lambda/ W$, yielding the ratio $L_g/L_k \propto W/\lambda$. On the other hand, for $S \ll W$, the geometric inductance scales with geometry as $L_g \rightarrow \mu_0 S/W$, yielding $L_g/L_k \propto S/\lambda$. Over the entire parameter range, we expect $L_g/L_k$, and thus $Q_{\textrm{QP}}$, to increase with both $S$ and $W$.

To experimentally investigate the dependence of resonator quality factor on geometry, we have characterized a series of hanger-style quarter-wavelength CPW resonators fabricated from 95 nm thick Nb films \mbox{($T_c =$ 8.8 K; RRR = 3.6)} sputtered on single crystal Al$_2$O$_3$ (0001) substrates. The traces were defined via optical lithography and a chlorine-based reactive ion etch (RIE). Each 6.25$\times$6.25 mm$^2$ chip accommodated six resonators multiplexed in frequency over a bandwidth of 400 MHz centered at 5 GHz [see Fig.~\ref{Fig1}(b)]. The resonators were capacitively coupled to the feed line via an elbow coupler 500 $\mu$m in length, yielding a coupling capacitance of $\sim$ 5 fF. Center trace widths of 5, 10, 20, 30, and  50 $\mu$m were studied. For each center trace width, the CPW gaps ranged from 1-30 $\mu$m. Devices were cooled to 4.2~K in an LHe dip probe, and transmission across the resonators was measured to extract $Q_{\textrm{QP}}$. In total, 150 resonators were characterized.

The forward scattering parameter of a quarter wave shunt resonator is well described by\cite{mazinThesis}
\begin{equation}\label{mazinS21}
S_{21} = \frac{S_{\text{min}} + 2iQ\text{$\delta$}x}{1 + 2iQ\text{$\delta$}x}.
\end{equation}
\begin{figure}[t]
\centering
	\includegraphics[width=.49\textwidth]{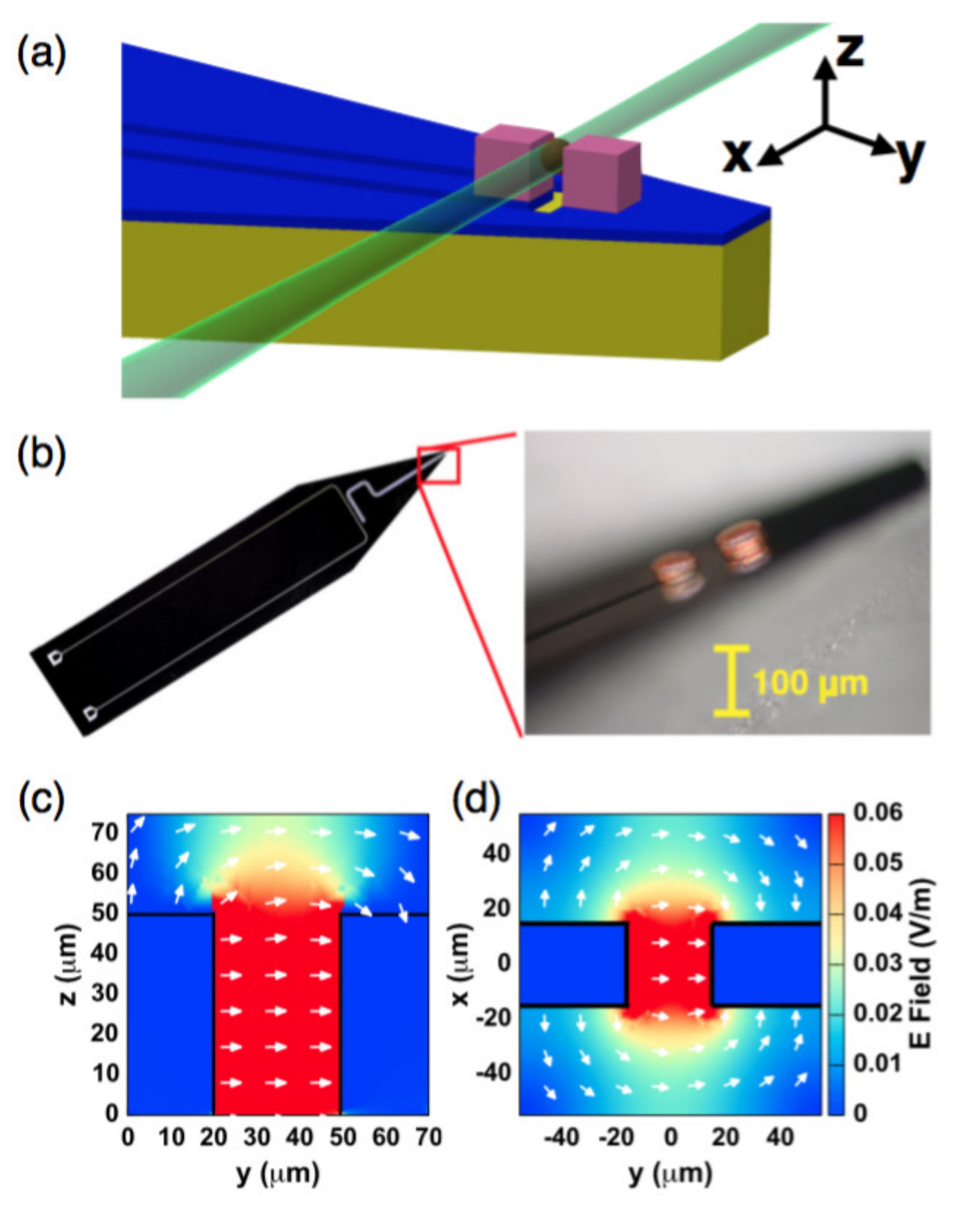} \hfill \\[-2ex]
	\caption{(Color Online) (a) Schematic of proposed superconductor--atom interface. (b) Micrograph of the superconducting chip and interaction region. Cu electrodes were plated to a height $\sim 50 \, \mu$m to facilitate coupling to trapped atoms tens of $\mu$m from the chip surface. (c-d) Profile and overhead view of the CPW microwave electric field at the gap capacitor for $V = 2 \, \mu$V and for an electrode spacing of $30 \, \mu$m. Black lines indicate the edges of the electroplated structures. }
	\label{cpwCartoon}
\end{figure}
Here, $\delta x = (f - f_c)/f_c$ is the reduced frequency relative to the resonator center frequency $f_c$, $S_{\text{min}} = Q_c/(Q_i + Q_c)$ is the transmission on resonance, and $Q = \left(1/Q_i + 1/Q_c\right)^{-1}$ is the total quality factor of the resonator with internal and coupling quality factors $Q_i$ and $Q_c$, respectively. The resonator parameters were extracted from the data via least squares fitting of Eq.~\eqref{mazinS21}. In Fig.~\ref{Fig1}(c), we plot the fitted $Q_i$ for all resonators along with the corresponding predictions from Eq.~\eqref{Qcpw} as a function of resonator dimensions. We see that appropriate choice of resonator geometry enables an almost order-of-magnitude enhancement of $Q_{\textrm{QP}}$ compared to narrow-gap, narrow-linewidth devices commonly used in state-of-the-art circuit QED experiments. Independent tunneling measurements yield a superconducting energy gap for our Nb thin films of \mbox{$\Delta/e = 1.0$ mV;} for this value of the gap, the data are best fit with a zero-temperature penetration depth $\lambda_0~= 87 ~$nm, in good agreement with other measurements of $\lambda_0$ in Nb thin films \cite{Gubin:2005aa}. Given that each chip accommodates only 6 resonators, 5 different chips per center trace width were needed to fully characterize the dependence of quality factor on geometry. For each chip, the six resonators were multiplexed in frequency with a typical spacing of \mbox{50 MHz}. To achieve good agreement between theory and experiment, it was necessary to include the extracted resonator center frequencies in the calculation of the complex frequency-dependent conductivity $\sigma$. The discontinuous steps in both the experimental data and theoretical predictions are a result of slightly varying resonator center frequency across the chip.

Crucial to the implementation of our proposed hybrid superconductor--atom interface is the ability to strongly couple a single trapped Rydberg atom to a voltage antinode of the resonator \cite{Pritchard:2014aa}. For the standard thin-film CPW geometry, the electric fields fall off rapidly with distance from the chip surface; however, optically trapping a single atom within microns of the chip is not practical, due to the significant heat load on the chip from scattered trap light. To facilitate strong electric dipole coupling to a single Rydberg atom at a trap location that is tens of microns from the chip surface, we have developed a thick-film Cu electroplating process that enables incorporation of tall ($\sim$ 50 $\mu$m) trapping electrodes at the voltage antinode of the resonator [Fig.~\ref{cpwCartoon}(a)]. The chip design is shown in Fig. 2(b). Here, a quarter-wave resonator is inductively coupled to a microwave feed line; the Cu trapping structures are integrated at the voltage antinode of the resonator (figure inset). The elongated shape of the chip allows for the inclusion of necessary signal and ground wire bonds far from the CPW--atom interaction region. Additionally, the chip tapers to a width $ < 150 \, \mu$m at the end where the atom will be trapped, serving to minimize the amount of scattered laser light on the superconducting surface due to the finite Rayleigh range of the trapping beams. The dimensions of the resonator were chosen to be $W = 50 \,\mu$m and $S = 25 \, \mu$m with a thickness \mbox{$D = 190$ nm,} yielding a resonator impedance $Z_r \sim 50 \, \Omega$ and an expected quasiparticle-limited $Q$ at 4.2~K in excess of $10^4$. 

\begin{figure}[b]
\centering
	
	\vspace*{-5mm}\includegraphics[width=.49\textwidth]{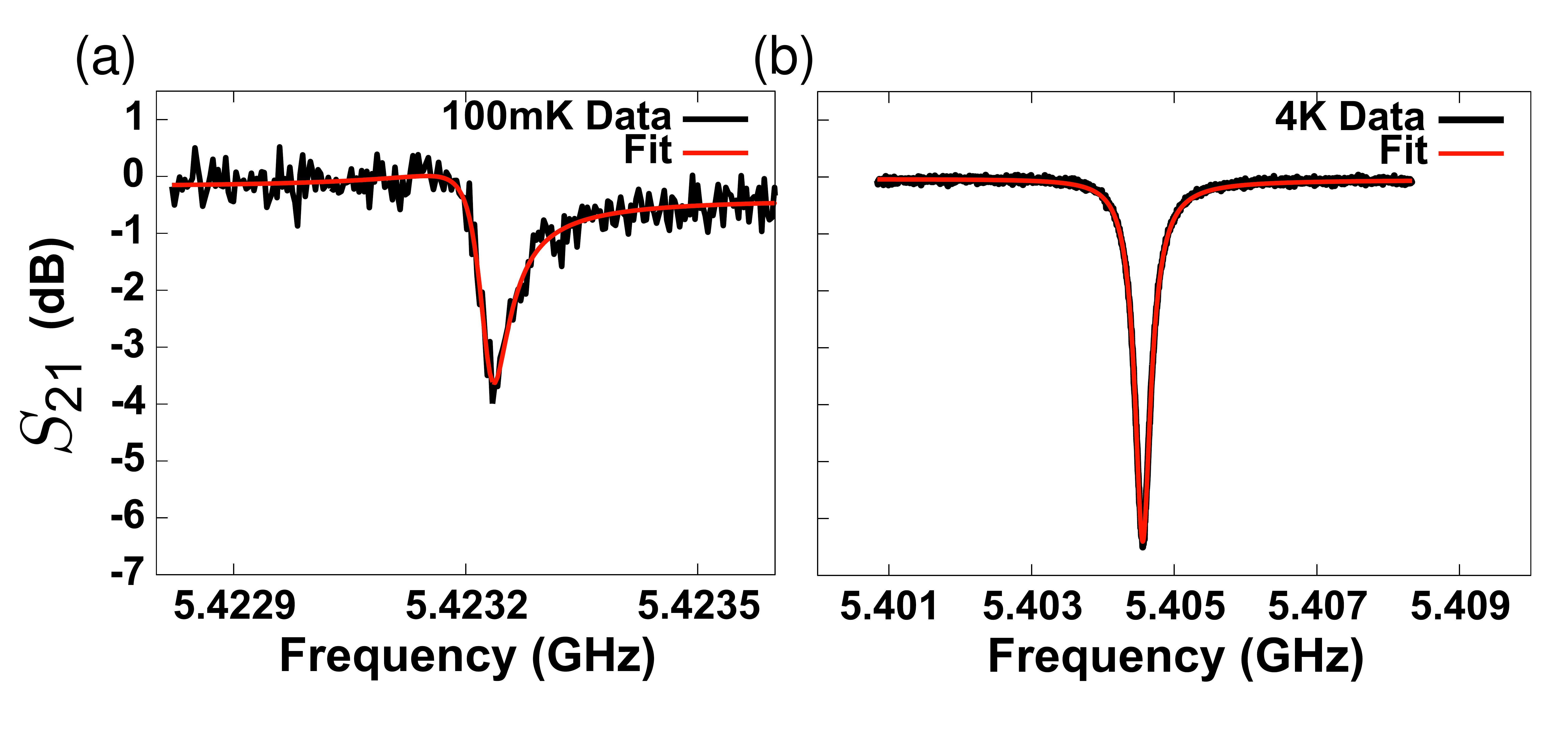} \hfill \\[-2ex]
	
	\caption{(Color Online) (a) Microwave transmission across an electroplated resonator measured at 100 mK. At single-photon power levels (shown), we find $Q_{\text{100 mK}} = 1.5\times10^5$. (b) Microwave transmission across a second electroplated resonator at 4.2~K. Note the different frequency scale. At this temperature, the internal quality factor is power-independent, with $Q_{4 \,\text{K}} = 3.0\times10^4$.}
	\label{epFig}
\end{figure}

 The Cu trapping structures were grown in a commercial sulphuric acid-based plating solution (Enthone Microfab SC) and pulse plated with a current density of 10 A/cm$^2$ across the wafer. Integrating the trapping structures on the Nb thin films required an intermediary adhesion layer of Ti/Pd grown by electron beam evaporation and patterned via lift off. Plating thicknesses of order 50 $\mu$m were achieved with a total charge transfer of 1500 A$\cdot$s. For the chosen resonator parameters, we expect zero-point electric fields between the trap structures $\langle V^2 \rangle^{1/2}= \sqrt{\hbar \omega_c / 2C_{ \text{CPW} } } = 2 \, \mu$V, where $\omega_c$~=~$2\pi \times 5.4$~GHz is the resonance frequency and $C_\text{CPW}=0.44$ pF. In ~Fig. 2(c-d)~ we show COMSOL simulations of the electric fields in the gap between the Cu trap structures. The simulations show that the electric field is uniform in magnitude and direction to roughly the height of the Cu structures with a peak field of $|\vec{E}| = 6.0 \times10^{-2}$ V/m.
 
In Fig.~\ref{epFig}, we show data from microwave scattering measurements on electroplated resonators characterized at both 100~mK and 4.2~K. At 100~mK, we find a low-power (single photon) quality factor $Q_i = 1.5\times 10^5$ and a high-power quality factor $Q_i=1.9\times 10^5$, in good agreement with measurements of similar resonators reported elsewhere \cite{Megrant:2012aa}. The quality factors at 4.2~K are power-independent with a value $Q_i = 3.0\times 10^4$. The factor of 2 increase in quality factor compared to the multiplexed resonators described in Fig. 1 is due to the $1/D$ dependence of $L_k$. It is clear from the data that the electrodeposition of Cu at the voltage antinode of the CPW resonator does not introduce additional loss. This result is not unexpected, since at this location there are no microwave currents that might couple to the lossy normal metal film of the trapping structures.

\begin{figure}[!]
\centering
	\vspace*{-11mm}\includegraphics[width=.5\textwidth]{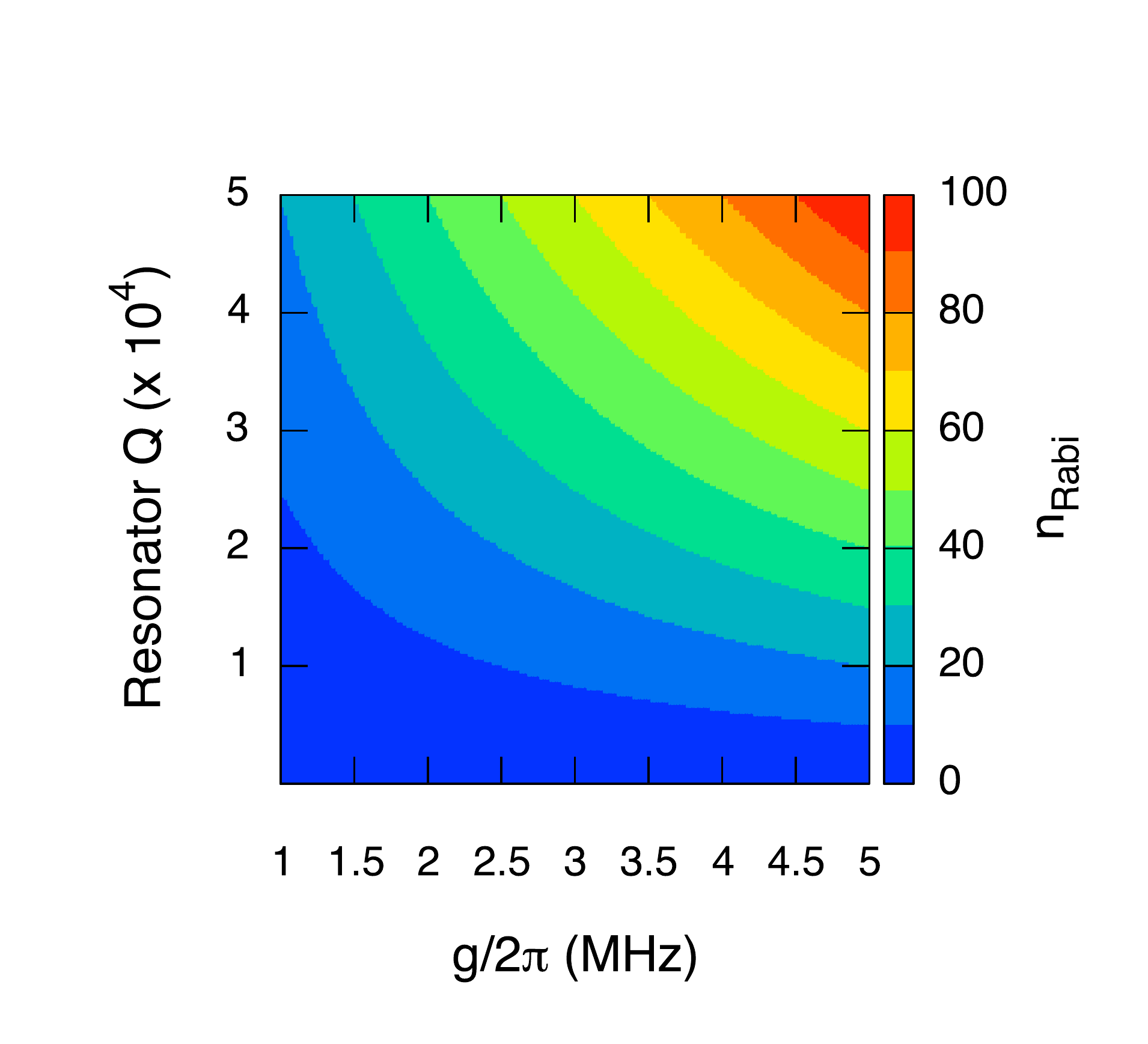}  \\[-5ex]
	\caption{(Color Online) Surface plot of $n_{\text{Rabi}}$. For the demonstrated resonator quality factor of $3.0 \times10^4$ and a coupling $g/2\pi = 3$ MHz, we expect to achieve $n_{\text{Rabi}} \approx 35$.}
	\label{nRabi}
\end{figure}

Our protocol relies on excitation of the single trapped atom to a high principal quantum number $n=88$ so that the Cs microwave transition $d_{rr'} = \braket{r|\text{\textbf{d}}|r'} =  \braket{88p_{3/2}, m=1/2|\text{\textbf{d}}|88s_{1/2},m=1/2} = \sqrt{1/6} \times 9210 \, ea_0$ with frequency ${\omega_{rr'} =  2\pi\times 5.406}$ GHz is near resonant with the CPW transition $\omega_c$ (additional fine tuning of the atomic transition to achieve resonance can be accomplished by dc Stark shifting the atomic levels). On resonance $\omega_{rr'} = \omega_c$ the atom and resonator will exchange a photon excitation at a rate equal to twice the vacuum Rabi frequency $\Omega = 2g$, where ${g = \text{\textbf{E}}\cdot \text{\textbf{d}}/\hbar}$. The number of superconductor--atom coherent oscillations within a photon lifetime is given by ${n_{\text{Rabi}} = 2g/(\gamma + \kappa)}$, where $\gamma, \, \kappa$ are the loss rates of the atom and resonator, respectively. The radiative decay times of $\ket{r}$ and $\ket{r'}$ are 1.9 ms and 750 $\mu$s \cite{Beterov:2009aa}, respectively, whereas the photon lifetime in the CPW cavity is ${\tau = Q/\omega_c \approx 1 \, \mu}$s. Accordingly, the number of superconductor--atom coherent oscillations reduces to $n_{\text{Rabi}} = 2gQ/\omega_c$. Fig.~\ref{nRabi} displays a surface plot of $n_{\text{Rabi}}$ as a function of resonator quality factor $Q$ and resonator--atom coupling rate $g$. Based on the calculated zero-point electric fields in the trapping region of the resonator, we anticipate a vacuum Rabi frequency ${g/2\pi \approx 3 \, \text{MHz}}$. Combined with demonstrated quality factors $Q = 3.0\times 10^4$, we should achieve $n_\text{Rabi} ~\approx 35$, placing the interaction securely in the strong coupling regime of cavity QED \cite{Wallraff:2004aa}. Moreover, this number compares favorably with that achieved in bulk Nb cavities and beams of Rydberg atoms\cite{Raimond:2001aa}.

More generally, we expect the resonators described here to serve as a fruitful test bed for a wide range of strong coupling superconductor--atom physics. They will enable investigation of the Purcell enhancement and suppression of atomic lifetimes as the atom is tuned into and out of resonance with the superconducting cavity in the trapping region. In addition, the strong dispersive interaction between the cavity and appropriately detuned Rydberg level should allow for a microwave-based quantum nondemolition measurement of the atomic state. Finally, the strongly-coupled single atom could be used as a local probe of stray electric fields due to surface adsorbates on the resonator chip \cite{Thiele:2015aa}.
\vspace{-0mm}

In conclusion, we have demonstrated that superconducting CPW resonator quality factors above $10^4$ are achievable at 4.2~K through appropriate engineering of the ratio of resonator geometric inductance to kinetic inductance. In addition, we have developed a method to increase the spatial extent of the zero-point electric fields at the resonator antinode without introducing additional loss. For the resonator parameters demonstrated here, strong coupling between a superconducting microwave mode and a single trapped Rydberg atom should be readily achievable.

Portions of this work were performed in the Wisconsin Center for Applied Microelectronics, a research core facility managed by the College of Engineering and supported by the University of Wisconsin-Madison. This work was supported by funding from NSF award PHY-1212448, ARO DURIP grant W911NF-15-1-0333 and ARO contract W911NF-16-1-0133.
%
%


\end{document}